\documentclass[12pt,onecolumn,oneside,draftcls]{IEEEtran}

\usepackage{graphicx}

\usepackage{graphicx,cite,amsmath,amssymb,hhline,subfigure}

\usepackage{epstopdf}

\usepackage[subnum]{cases}
\usepackage{amsmath}
\usepackage{multirow}
\usepackage{hyperref}
\usepackage{algorithm,algorithmic}

\begin{document}

\title{Analysis of Spectrum Occupancy Using Machine Learning Algorithms}
\author{
\bigskip
\medskip {\normalsize $\mbox{Freeha Azmat}^{}$, $\mbox{Yunfei Chen}^{}$, {\em Senior Member,
IEEE},
                    and
                   $\mbox{Nigel Stocks}^{}$ }}
\thispagestyle{empty} \setcounter{page}{0}
\maketitle

\renewcommand{\baselinestretch}{1.73} \normalsize

\begin{abstract}
In this paper, we analyze the spectrum occupancy using different machine learning techniques. Both supervised techniques (naive Bayesian classifier (NBC), decision trees (DT),
support vector machine (SVM), linear regression (LR)) and  unsupervised algorithm (hidden markov model (HMM)) are studied to find the best technique with the highest classification accuracy (CA). A detailed comparison of the supervised and unsupervised algorithms in terms of the computational time and classification accuracy is performed.   The classified occupancy status is further utilized to evaluate the probability of secondary user outage for the future time slots, which can be used by system designers to define spectrum allocation and spectrum sharing policies. Numerical results show that SVM is the best algorithm among all the supervised and unsupervised classifiers.
Based on this, we proposed  a new SVM algorithm by combining it with fire fly algorithm (FFA), which is shown to outperform all other algorithms. 
\end{abstract}


\newpage
\section{Introduction}
A cognitive radio network (CRN) is composed of two types of users, namely, the licensed primary users (PU's) and the unlicensed secondary users (SU's). The core idea behind CR is to allow unlicensed user's
access to the licensed bands in an opportunistic manner to avoid interference with the licensed users. To achieve this, a realistic understanding of the dynamic usage of the spectrum
is required. The spectrum measurement is an important step
towards the realistic understanding of the dynamic spectrum usage. Various spectrum measurement campaigns covering a wide range of frequencies have been performed \cite{chen}. These pectrum measurements studies have found significant amount of unused frequency bands
in the case of normal usage due to the  static spectrum regulations. This has led researchers to understand  the spectrum occupancy characteristics in depth for exploiting the free spectrum.

\subsection{Problem definition}
Many studies have been performed to understand the occupancy statistics.  For instance, the statistical and spectral occupation analysis of the measurements was presented in \cite{occupation} in order to study the traffic density in all frequency bands.  In \cite{radio}, autoregressive model was used to predict the radio resource availability using occupancy measurements in order to achieve uninterrupted data transmission of secondary users. In \cite{predictive}, the occupancy statistics were utilized to select the best channels for control and data transmission purposes, so that less time is required for switching transmission from one channel to the other for the case when the PU appears. Further, In \cite{power}, \cite{duration}, the bandwidth efficiency was maximized by controlling the transmission power of cognitive radio using spectrum occupancy measurements. 

In \cite{statistics}, different time series models were used to categorize specific occupancy patterns in the spectrum measurements. All of the aforementioned works have evaluated the spectrum occupancy models by using conventional probabilistic or statistical tools. These tools are often limited due to assumptions required to derive their theories. For example, one has to determine whether the value is random variable or a random process in order to use either probabilistic and statistical tools. On the other hand, machine learning (ML) is a very powerful tool that has received increasing attention recently  \cite{ML}. The machine learning algorithms are often heuristic, as they don't have any prerequisites or assumptions on data. As a result, in many cases, they provide higher accuracy than conventional probabilistic and statistical tools.  There are very few works on the use of ML in spectrum occupancy. For example, the ML works related to CR in \cite{ml1}- \cite{linear} discussed cooperative spectrum sensing and spectrum occupancy variation. However, in this paper, we aim to provide a comprehensive investigation on the use of  ML for analyzing spectrum occupancy. The motivation is that different ML algorithms are often suitable for different types of data. Thus, one needs to try different ML algorithms in order to find the one that suits the spectrum data best, not just one ML algorithm.
 \subsection{ Contributions} 
The contributions are listed as follows:\\
1. We propose the use of ML algorithms in spectrum occupancy study.  Both supervised and unsupervised algorithms are used. The machine learning techniques are advantageous because they are capable
of implicitly learning the surrounding environment and are much more adaptive compared with
the traditional spectrum occupancy models. They can describe more optimized decision regions on
feature space than other approaches. In \cite{ml1} and \cite{ml2}, ML was used for cooperative spectrum sensing. However we use ML for spectrum occupancy modelling that may be used in all CR operations, including spectrum management, spectrum decision and spectrum sensing. In \cite{review}, authors have discussed call-based modelling for analyzing the spectrum usage of the dataset collected from the cellular network operator. Further, they have shown that random walk process can be used for modeling aggregate cell capacity. However, we use ML to model spectrum occupancy in time slots for all important bands. \\
2.  We have utilized four supervised algorithms, naive Bayesian classifier (NBC), decision trees (DT), support vector machine (SVM), linear regression (LR), and one unsupervised algorithm, hidden markov model (HMM), to classify the occupancy status of time slots. The classified occupancy status is further utilized for evaluating the probability of SU outage. In \cite{pred}, HMM was used to predict the channel status. Our supervised algorithms and modified HMM all perform better than HMM. In \cite{linear}, LR was used  to investigate the spectrum occupancy variation in time and frequency. Our approach outperforms LR as well.
                               
3. We propose a new technique that combines SVM with fire fly algorithm (FFA) that outperforms all supervised and unsupervised algorithms.

The rest of the paper is organized as follows: Section II explains the system model, followed by the detailed explanation of classifiers in Section III. The numerical results and discussion are presented in Section IV.

\section{System Model}
\subsection{Measurement setup and data}
We have measured the data from 880 MHz to 2500 MHz containing  eight  main radio frequency bands for approximately four months (6th Feb-18th June 2013) at the University of Warwick using radiometer. The eight bands are: 880-915 MHz, 925-960 MHz, 1900-1920 MHz, 1920-1980 MHz, 1710-1785 MHz, 1805-1880 MHz, 2110-2170 MHz and 2400-2500 MHz. The number of the frequency bins in each band varies. For example, the band 925-960 MHz contains 192 frequency bins, each occupying a bandwidth of 0.18 MHz, while the band 1710-1785 MHz contains 448 frequency bins, each occupying a bandwidth of 0.167 MHz. The data is arranged in a two dimensional matrix ($t_i,f_j$) for each band; where each row $t_i$ represents the measured data at different frequencies in one minute while each column $f_j$ represents the data at different time instants of each frequency bin. As we have measured the data for four months which constitute 131 days (188917 minutes), the numbers of rows are 188917 while the number of columns varies according to the number of the frequency bins in a particular band.

\subsection{SU Model}

In a network of licensed users, SU is allowed to access the licensed band without causing any harmful interference to the PU. Let $i$ denote the time slot and $j$ denote the frequency bin, where $i=1,2,..n$, $j=1,2,...k$, $n$ represents the total number of time slots and $k$ represents the total number of frequency bins. Using energy detection \cite{energy}, if $y^i(j)$ is the sample sensed at the $i^{th}$ time slot in the $j^{th}$ frequency bin. One has
\begin{subequations}
\begin{equation}
\label{e2}
\hspace {10mm} y^i(j)=x^i(j)+w^i(j)
\end{equation}
\begin{equation}
\label{e2rr}
or \hspace {10mm} y^i(j)=w^i(j)
\end{equation}
\end{subequations}
where $x^i(j)$ represents the received PU signal and $w^i(j)$ represents the additive white Gaussian noise (AWGN) with zero mean and variance $\sigma_{w}^2$. Each sample is compared with a threshold ($\gamma$). The selection of $\gamma$ is very important because small values of $\gamma$ will cause false alarms while large values will miss spectrum opportunities. The computation of $\gamma$ was explained in \cite{utility}. In our approach, the threshold is dynamic and its selection is explained in Section IV-B. The spectrum status is given as

\[
S^i(j)=
\begin{cases}
\label{2}
1,&  y^i(j)>\gamma\\
0,&  y^i(j)<\gamma.
\end{cases}
\] 
 The occupancy for the $ith$ time slot for all $k$ frequency bins is defined as
\begin{equation}
\label{e4}
OC^i=\frac{\sum_{j=1}^k S^i(j)}{k}
\end{equation}

%


\begin{figure}
\begin{center}
\includegraphics[width=7in, draft=false]{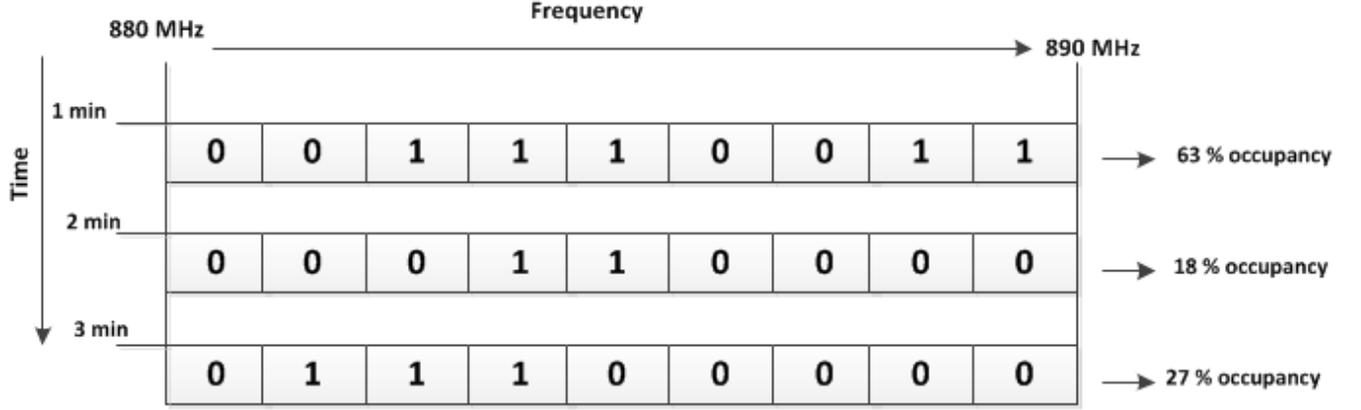}
\caption{Occupancy for different time slots in the band.}
\label{f1}
\end{center}
\end{figure}
For example, a three minutes interval for the band 880 - 890 MHz having 9 frequency bins is shown in Fig.\ref{f1}, where each bin occupies 1MHz. For each frequency bin, $S^i(j)$ is decided. Once $S^i(j)$ is evaluated, the occupancy $OC^i$ is calculated using (\ref{e4}). It is observed that more frequency bins are occupied for the first minute than for the second and third minutes so that it has less chance for SU to transmit. Following the discussion above, we need to set the criteria for quantifying this chance based on the occupancies.
\subsection{PU Model}
As per our approach, the status of PU ($P^i$) for each $i^{th}$ time slot can be decided using the following rules:\\
\[
P^i=
\begin{cases}
1,&  OC^i>U_{oc} \hspace{3mm} (Condition \hspace{1mm} 1)\\
1,&  L_{oc}<=OC^i<=U_{oc} \hspace{3mm} AND \hspace{3mm} con^i<B \hspace{3mm} (Condition\hspace{1mm}  2)\\
0,&  L_{oc}<=OC^i<=U_{oc} \hspace{3mm}   AND \hspace{3mm} con^i>=B \hspace{3mm} (Condition\hspace{1mm}  3)\\
0,&  OC^i<L_{oc} \hspace{3mm} (Condition\hspace{1mm}  4)\\\
\end{cases}
\] 

where $U_{oc}$ and $L_{oc}$ represents the maximum and minimum values of occupancy for all $n$ time slots, $con^i$ represents the number of consecutive free frequency bins in each $ith$ time slot and $B$ represents the maximum value of $con^i$, when PU is considered present. Each condition is explained as follows:

1. Condition 1 and Condition 4:
The values of $U_{oc}$ and $L_{oc}$ vary with the frequency band, the day and the threshold. Our test show that $U_{oc}$ should not be less than 75$\%$ and $L_{oc}$ should not be greater than 40$\%$. For fixed frequency band and day, we have evaluated $U_{oc}$ and $L_{oc}$ for different thresholds  in Section IV-B.  In order to guarantee PU protection and ensure SU transmission when the values of $OC^i$  lie in the range between $L_{oc}$ and $U_{oc}$, further criterion is applied.
 
2. Condition 2 and Condition 3:
When $L_{oc}<=OC^i<=U_{oc}$, it is difficult to apply condition 1 and condition 4. So we evaluate $con^i$ for each time slot. If $con^i>B$ for $L_{oc}<=OC^i<=U_{oc}$ , there exists at least $B$ consecutive free frequency bins in $ith$ time slot; thus SU can transmit and vice versa when $con^i>B$. The value of B is selected to provide PU protection. This will be explained in Section IV-B. 

\subsection{ Machine Learning Framework for SU and PU Model}
ML constructs a classifier to map $\textbf{S}^i$
to $P^i$, where $\textbf{S}^i=[S^i(1),S^i(2),..S^i(k)]$ represents the feature vector and $P^i$ is the corresponding response to the feature vector. There are two steps for constructing a classifier:
\subsubsection{Training}  Let
$\textbf{S}^i_{train}  = [S^i(1)_{train},S^i(2)_{train} . . . , S^{i}(k)_{train}]^T$ denote the training spectrum status and $P^i_{train}$ represent the training PU status for the $ith$ time slot respectively, where $i=1,2,..n1$ and $n_1$ represents the number of training time slots fed into the classifier.
\subsubsection{Testing}
Once the classifier is successfully trained, it is ready to
receive the test vector for classification. Let $\textbf{S}^i_{test}  = [S^i(1)_{test},S^i(2)_{test} . . . , S^{i}(k)_{test}]^T$ denote the testing spectrum status and $P^i_{test}$ represent the testing PU status for the $ith$ time slot respectively, where $i=n1+1,n1+2,..n2$ and $n_2$ represents the length of testing sequence. It is assumed that $n=n_1+n_2$. For our proposed approach, the matrix of size $n*k$ is divided into $15\%$ training data matrix of size $n_1*k$ and $85\%$ testing data matrix of size $n_2*k$. The value $P^i_{test}$ is not used during the testing but as a reference for computing the classification error.
\subsubsection{Classification Accuracy (CA)}  
Let $P^i_{eval}$  denote the PU status determined by the classifier for the $ith$ time slot. The classifier categorizes the testing vector  $\textbf{S}^i_{test}$ as 'occupied class' (i.e., $P^i_{eval} = 1$) or 'unoccupied class' (i.e., $P^i_{eval} = 0$). Therefore, the PU status is correctly determined, when $P^i_{eval}$= $P^i_{test}$, giving $CA^i=1$. The misdetection occurs, when $P^i_{eval} = 0$ and $P^i_{test} = 1$ while false alarm occurs, when $P^i_{eval} = 1$ and $P^i_{test} = 0$, giving $CA^i=0$.

\subsection{Probability of SU outage}
Let $\textbf{P}^i_{eval}$  be a vector of length $((n_2-n_1)+1)$ evaluated by each classifier, and $P^i_{eval}$ represent the presence/absence of PU for the $i^{th}$ time slot. When  $P^i_{eval}=0$,  SU is allowed to utilize the $i^{th}$ time slot. Define $out_{su}$ as the minimum value of consecutive free time slots required by SU for transmission. SU outage occurs, when SU cannot find $out_{su}$ consecutive free time slots in a vector $\textbf{P}^i_{eval}$ of length $((n_2-n_1)+1)$. The probability of SU outage is given by

\begin{subequations}
\begin{equation}
\label{e11a}
P(SU_{outage})=1-P(SU_{transmit})
\end{equation}
where
\begin{equation}
\label{e11ab}
P(SU_{transmit})=\sum_{c=1}^{C} P(FB_c)
\end{equation}
where $FB_c$ represents the block of free consecutive time slots of length $out_{su}$, $c=\{1,2,..C\}$ and $C$ represents the total number of free blocks present in $P^i_{eval}$. The probability for a free block starting at index, say $r$ in $P^i_{eval}$ is evaluated using the following equation 
\begin{equation}
P(FB_c)=\prod_{i=r}^{r+out_{su}} OC^i. 
\end{equation}
\end{subequations}

\section{Proposed Algorithms}
In the proposed approach, five machine learning algorithms are utilized to predict the future PU status using the occupancy data, which is a function of time, frequency and threshold. Among them, four are supervised learning algorithms: NBC, DT, SVM and LR, while one is an unsupervised algorithm, HMM. The motivation to use five different algorithms is to find the best  machine learning algorithm as they have different characteristics. 
\subsection{Naive Bayesian Classifier}
A Naive Bayesian classifier is a generative model based on the Bayes theorem. It is also called 'independent feature model' because it does not take dependency of features into account. The feature vector for the $ith$ time slot in our model contains all the samples which are independent of each other, since every feature represents a specific frequency bin. For example, the status vector of the $ith$ time slot is given as $\textbf{S}^i={S^i(1), S^i(1), S^i(2),..,S^i(k)}$, where $S^i(1)$ is independent from $S^i(2)$. However, the response variable in our approach i.e. PU status ($P^i$) is a dependent variable which is affected by each frequency bin. As our features are independent, so we will use NBC for classification. The probability of  $\textbf{S}^i$ belonging to the class $P^i$ evaluated using the Bayes theorem is formally defined as \cite{nbapple}

 \begin{equation}
\label{e7}
p(P^i,\textbf{S}^i)=p(P^i)*p(\textbf{S}^i|P^i).
\end{equation}

when $P^i=0$, $\textbf{S}^i$ will be classified as 'idle' class, while when $P^i=1$, $\textbf{S}^i$ will be classified as 'occupied' class. The goal is to find the class with the largest posterior probability in the classification phase. The classification rule is given as
 \begin{equation}
\label{e8}
classify(\hat{\textbf{S}^i})= argmax_{\textbf{S}^i}\{p(P^i,(\hat{\textbf{S}^i})\}
\end{equation}
where $\hat{\textbf{S}^i}= \{\hat{S^i(1)}, \hat{S^i(2)}...\hat{S^i(k)}\}$. NBC is sensitive to the choice of kernel and the prior probability distribution of classes. This  will be explained in Section IV-B.

\subsection{Decision Trees}
Decision tree builds classification or regression models in the form of a tree structure. The decision trees used in this approach are classification trees whose leaf represents the class labels. Unlike NBC, it can handle feature interactions and dependencies. In DT, the decision is made on each internal node which is used as a basis for dividing the data into two subsets while leaf nodes represent the class labels (in the case of classification trees) or the real numbers (in the case of regression trees). Data come in the form

\begin{equation}
\label{e9}
(\textbf{S}^i,P^i)=(S^i(1),S^i(2),S^i(3)..,S^i(k),P^i).
\end{equation}
where $P^i$ is the dependent variable representing the class label of $ith$ time slot. The class labels $P^i$ are assigned by calculating the entropy of the feature, as \cite{trees}
\begin{equation}
\label{e10}
Entropy(t)= -\sum_{id=0}^{Z}p(id|t)\hspace{1mm}log_2p(id|t).
\end{equation}
Where $p(id|t)$ denote the fraction of records belonging to class $id$ at a given node $t$ and $Z$ represents the total number of classes. In our approach, $Z=1$. The smaller entropy implies that all records belong to the same class. It will be discussed in Section IV-C on how fraction of records per node affects the classification accuracy of DT. 
\subsection{Support Vector Machines}
SVM is a discriminative classifier with high accuracy. Unlike DT, it prevents over-fitting and can be used for online learning \cite{puni}. There are two types of classifiers in SVM: linear SVM for separable data and non-linear SVM for non-separable data. The linear classifier is used here. The training feature and response vectors can be represented as $D={(P^i,\textbf{S}^i)}$  where $P^i\in\{0,1\}$ . The two classes are separated by defining a random  division line $H$ represented as $d.\textbf{S}^i+b=\rho$, where $d$ and $b$ represent the weighting vector and bias, respectively, while $\rho$ represents the constant for dividing two hyper planes. The maximum-margin hyper planes that divide the points having $P^i=1$ from those $P^i=0$ are given as: 
\begin{subequations}
\begin{equation}
\label{e11a}
P^i=+1 \hspace{5mm}when \hspace{5mm} d.\textbf{S}^i+b>\rho\hspace{10mm}  (Occupied \hspace{1mm} Class)
\end{equation}
\begin{equation}
\label{e11b}
P^i=0 \hspace{5mm}when \hspace{5mm}
 d.\textbf{S}^i-b<\rho \hspace{10mm}  (Idle \hspace{1mm} Class)
\end{equation}
\end{subequations}
The separation between two hyper planes is margin, controlled by the parameter called box constraint $Box_{ct}$. We have evalauted the optimal value of $Box_{ct}$  using a bio-inspired technique i.e. FFA in our approach.
\subsection{SVM with Fire Fly Algorithm}
In FFA, let $X$ be a group of fire flies, $X=[l_{1},l_{2},..l_{X}]$, initially located at specific positions $a_{X}=[a_{l_{1}},a_{l_{2}},..a_{l_{X}}]$. Each fire fly moves and tries find a brighter fire fly, which has more light intensity than its own. The objective function $f(x)$ used for evaluating the brightness of the fire fly in our approach is the classification accuracy i. e. $f(x)=CA(a_{X})$. When a fire fly, say $l_{1}$ finds another brighter fire fly $l_{2}$ at another location having more intensity compared to its own, it tends to move towards fire fly $l_{2}$. The change in position is determined as \cite{fire}
\begin{equation}
\label{e12}
a_{l_{1}}^{v+1}=a_{l_{1}}^v+\beta_0e^{-\psi_{l_{1}l_{2}}rd_{l_{1}l_{2}}^2}(a_{l_{2}}^v-a_{l_{1}}^v)+\alpha(rand-0.5)
\end{equation}
where $v$ represents the number of iterations, $a_{l_{1}}$ and $a_{l_{2}}$ represents the position of fire fly $l_{1}$ and $l_{2}$ respectively, $\alpha$, $\beta_0$ and $\psi_{l_{1}l_{2}}$ are constants and $rand$ is a uniformly distributed random number. For our approach, the starting positions of the $X$ fire flies are initialized, while the position of each fire fly represents the value of box constraints $Box_{ct}$. 

\subsection{Linear Regression}
The flexibility of linear regression to include mixture of various features in different dimensions e. g. space, frequency, time and threshold as a linear combination is the main motivation of using it for modeling in this approach. The linear regression model for our approach is given by:
\begin{equation}
\label{e13}
P^i=e_0+e_1S^i(1)+e_2S^i(2)+...+e_kS^i(k)= e_0 +\sum_{j=1}^{k} e_jS^i(j).
\end{equation}
where the class label $P^i$ is represented as a linear combination of parameters $e_1, e_2,…,e_k$ and features ($S^i(1), S^i(2),.., S^i(k)$) in the  $ith$ time slot. The stepwise-linear regression is used in this approach.  In each step, the optimal term based on the value of defined 'criterion' is selected. The 'criterion' can be set as the sum of squares error (SSE), deviance, akaike information criterion (AIC), Bayesian information criterion (BIC) or R-squared etc. SSE is used in this approach. The small values of SSE are encouraged for a good model. It is observed from (\ref{e13}), that the computational time for evaluating the response of the model linearly increases with the number of frequency bins/ predictors involved. So we need to select an appropriate number of predictors for linear regression.

\subsection{Hidden Markov Models}
It is an unsupervised algorithm for modeling the time series data. The motivation to use the unsupervised algorithm is that it does not need the training phase. In HMM, the sequence of states can be recovered by an analysis of the sequence of observations. The set of states and observations are represented by $U$ and $G$ given as $U=(u_1,u_2,...u_N)$, $G=(g_1,g_2,...g_M)$, where $u_1$ and $u_2$ represent the states when $P^i=0$ and $P^i=1$, respectively. The observations $g_1$ and $g_2$ represent the value of $OC^i$ corresponding to each $P^i$. HMM is defined as
\begin{equation}
\label{e14}
\lambda=(C_h,D_h,\pi)
\end{equation}
where the transition array $C_h$ is the probability of switching from state $u_{1}$ to state $u_{2}$ given as \cite{hmm}, $C_h=[c_{12}]=P(q_t=u_{2}|q_{t-1}=u_1)$. The $D_h$ is the probability of observation $g_{1}$ being produced from state, $D_h=[d_{{1},{2}}]=P(o_t=g_{{1},{2}}|q_{t}=u_{2})$ and $\pi$ is the initial probability array, $\pi=P(q_1=u_2)$.

HMM has two main steps. In the first step, the sequence of observations $O=(o_1,o_2,...o_T)$, transition probability matrix $C_h$ and emission probability matrix $D_h$ are utilized to find the probability of observations $O$ given hmm model $\lambda$ given in (\cite{hmm}, Eq.13) as, $P(O|\lambda)=\sum_QP(O|Q,\lambda)P(Q|\lambda)$, where $Q=(q_1,q_2,...q_T)$ and $P(O|Q,\lambda)=\prod_{t=1}^TP(o_t|q_t,\lambda)=g_{q_1}(o_1)*g_{q_2}(o_2)..g_{q_T}(o_T)$. The probability of the state sequence is given as $P(Q|\lambda)=\pi_{q_1}c_{q_1q_2}c_{q_2q_3}...c_{q_{T-1}q_T}$.
In the second step, the hidden state sequence, that is most likely to have produced an observation is decoded using the viterbi algorithm. The most likely sequence of states $Q_L$ generated using the viterbi algorithm is matched with the expected fixed state sequence $Q$ to compute classification accuracy. HMM can be also be supervised by adding two extra steps as \\ 
\textbf{Step(a)}: Use the initial guesses of $C_h$ and $D_h$ to compute $Q$ and $O$, that are used for computing $P(O|\lambda)$ in forward algorithm\\
\textbf{Step(b)}: Use $O$, $D_h$ and $C_h$ in Step(a) to estimate the transition probability matrix $C_{h'}$ and emission probability matrix $D_{h'}$ using maximum likelihood estimation \cite{hmm1}.\\
The $C_{h'}$ and $D_{h'}$ collectively form the estimated HMM model ($\lambda_e$)  that can be further used for evaluating $P(O|\lambda)$ and $Q_L$ using the forward algorithm and the Viterbi algorithm respectively.

\section{Numerical Results and Discussion}
In order to analyze the occupancy of the eight bands, the statistics of data in all bands from 880 to 2500 MHz are presented in Section IV-A. The classification criteria are explained in Section IV-B. The selection of the best parameters for each model using the classification criteria are discussed in Section IV-C. The classification models with the optimal parameters are compared to find the best classifier in terms of the CA, defined as $CA=\frac{\text{No. of correct classfications}}{\text{Total number of test samples}}$

\subsection{Statistics of Data}
\begin{figure}
\begin{center}
\includegraphics[width=4in, draft=false]{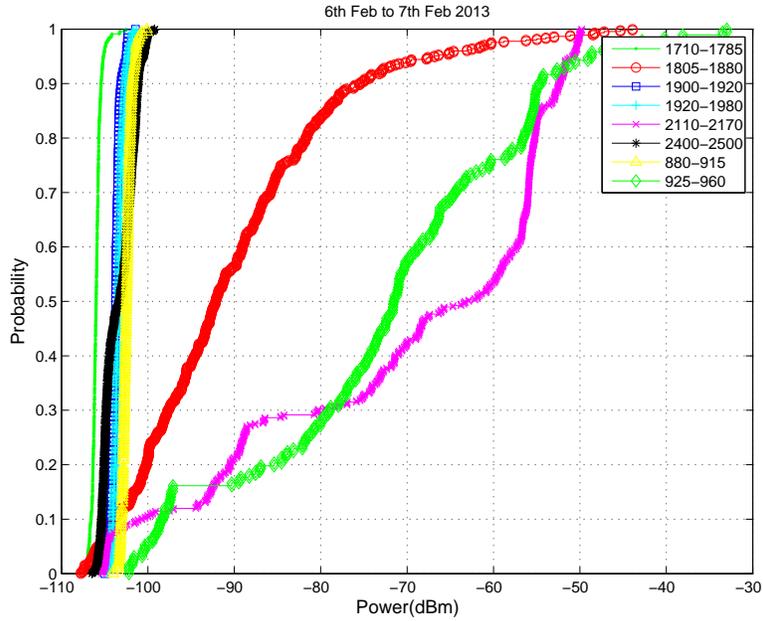}
\caption{The CDFs for the eight bands between 880-2500 MHz.}
\label{f4}
\end{center}
\end{figure}

The CDF plot is shown in Fig.\ref{f4} which  gives the summarized view of all power ranges for the eight bands. It can be observed from Fig.\ref{f4} that the eight bands can be categorized into two main groups. Group A contains those bands that have wide power ranges between -110 dBm to -30 dBm including 1805-1800 MHz, 1710-1785 MHz and 2110-2170 MHz. Group B has five bands: 925-960 MHz, 880-915 MHz, 2400-2500 MHz, 1920-1980 MHz and 1900-1920 MHz that have power ranges between -110 dBm and -100 dBm. Thus, Group A bands have larger standard deviation than Group B bands.  Next we discuss the effects of two main parameters (frequency and threshold) on occupancy. 
\subsubsection{Occupancy Vs Threshold}
\begin{figure} 
   \begin{center}
 
  \includegraphics[width=4.5in]{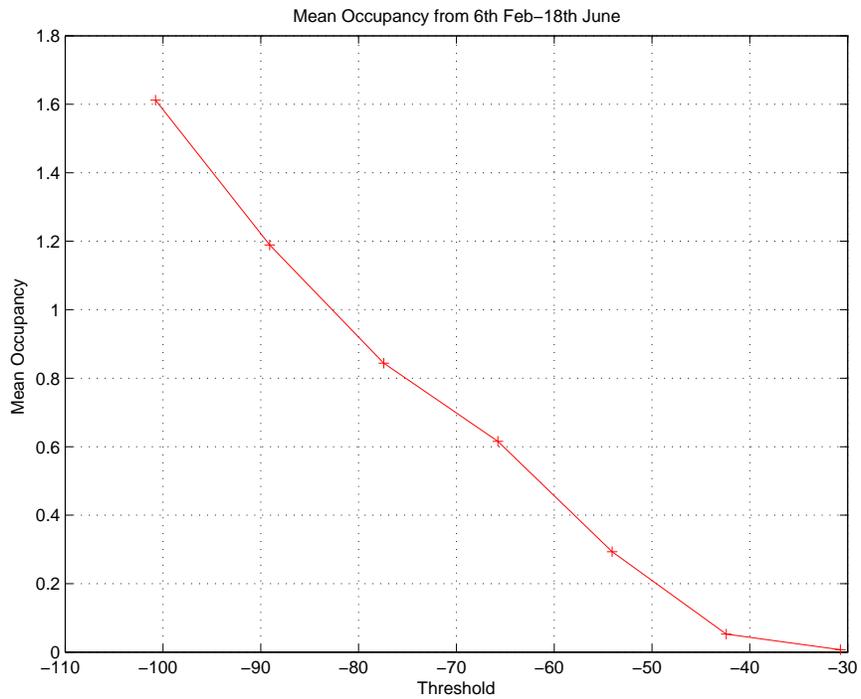}

    \end{center}

   \caption{Occupancy VS threshold for Band 925-960 MHz}
\label{f8}
\end{figure}

The threshold selection is an important task for analyzing the occupancy of each time slot. We took the minimum and the maximum value of power for each frequency band and tested seven values of thresholds in this range. Each band is analyzed separately for the seven values of the threshold using the four months data. Due to limited space, only 925-960 MHz is given in Fig.\ref{f8}. It is observed that occupancy monotonically decreases when the value of threshold increases. These results have proved that larger value of threshold will classify less samples as occupied.

\subsubsection{Occupancy Vs Frequency}
\begin{figure}

  \subfigure[\label{a}]{\includegraphics[width=4.5in]{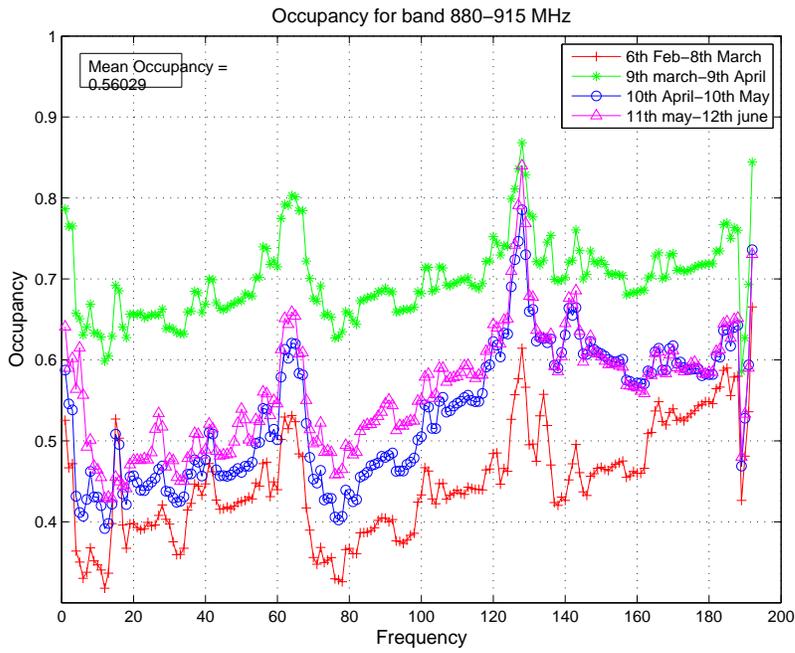}}

    \subfigure[\label{b}]{\includegraphics[width=4.5in]{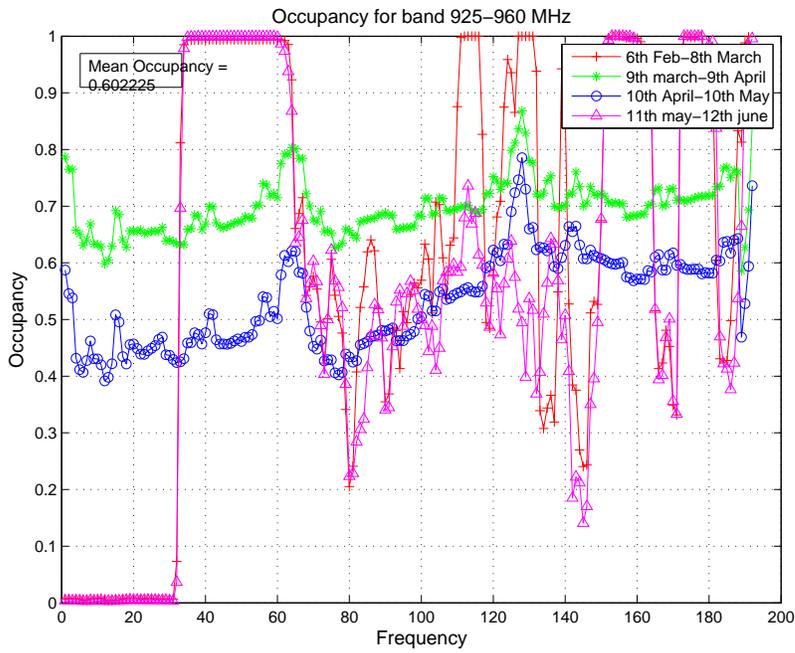}}

   \caption{Occupancy VS spectrum frequency for (a) Band 880-915 MHz (b) 925-960 MHz.}
\label{f9}
\end{figure}

The relationship between occupancy and frequency is analyzed by computing the occupancy of the $jth$ bin individually. Eq.(\ref{e4}) can be modified for computing the  occupancy of the $jth$ frequency bin ($OC^j=\frac{\sum_{i=1}^n S^i(j)}{n}$).
We have found in Fig.\ref{f9} a unique periodicity
in some bands. We found that four bands can be categorized as the periodic group bands: 880-915 MHz, 1710-1785 MHz, 2110-2170 MHz and 2400-2500 MHz bands. The bands 925-960 MHz, 1805-1880 MHz, 1920-1980 MHz and 2110-2170 MHz do not have this property.

The periodicity may be caused by the usage pattern. For instance, the periodicity in each band lies in their uplink/downlink usage pattern. For instance, the bands 1710-1785 MHz and 1900-1920 MHz are uplinks,  while the aperiodic bands 1805-1880 MHz and 1920-1980 MHz are downlinks. The uplink transmits data from the mobile user to base station so that its activity is completely determined by mobile users's periodic usage pattern. On the other hand, the downlink transmits the data from base station to the mobile user so that its activity is also affected by control and broadcast channels, making it less or non periodic. 


\subsection{ Classification Criteria}
\begin{figure}
\begin{center}
\includegraphics[width=7in, draft=false]{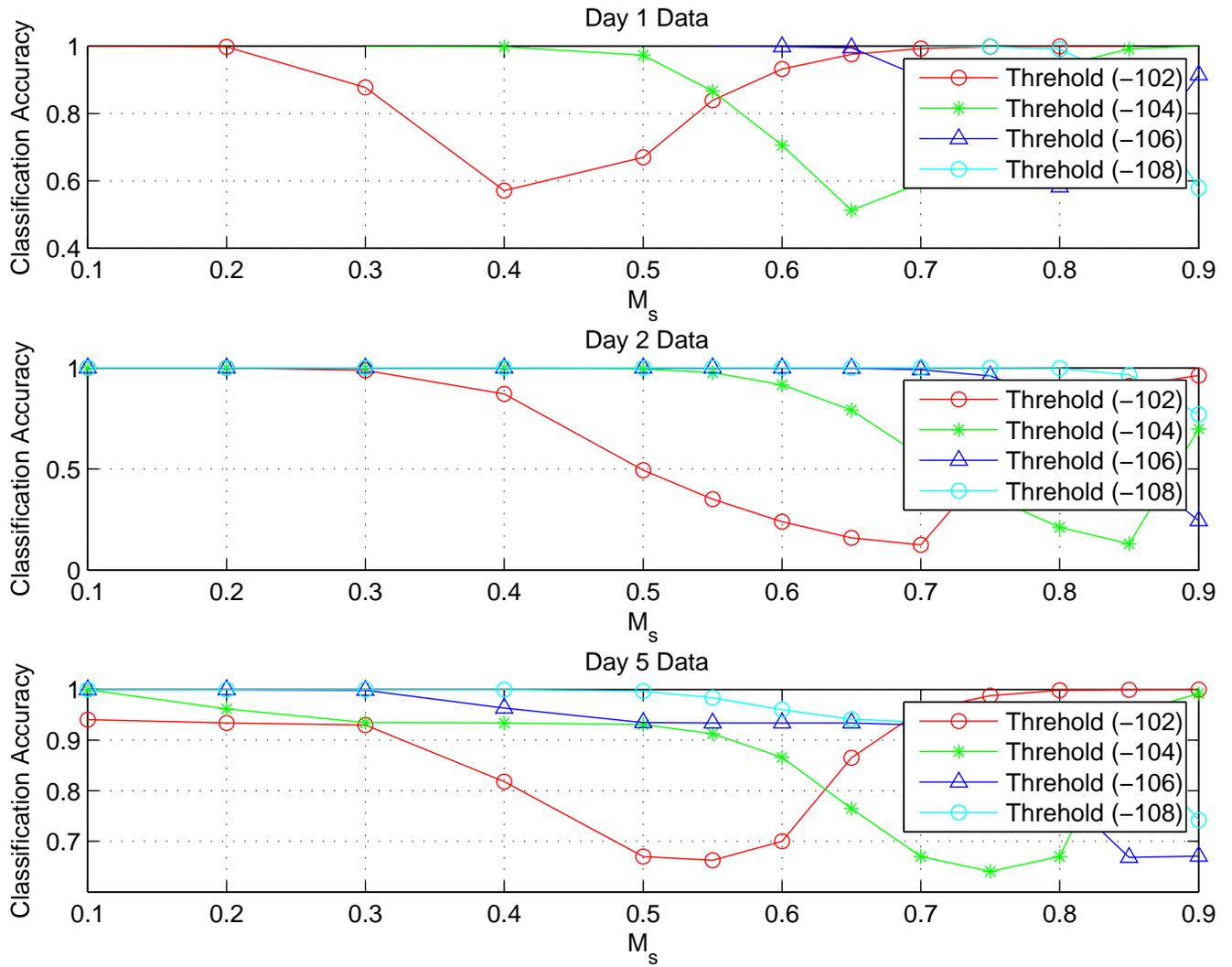}
\caption{Selection of optimal threshold ($\gamma$) and optimal splitting range ($[U_{oc}, L_{oc}])$ for determining the classification criteria of three days data.}
\label{f11}
\end{center}
\end{figure}

This subsection studies the choice of $U_{oc}$, $L_{oc}$, $con^i$ and B in Section II-C as shown in Fig \ref{f11}. We have utilized Day1 (1-1440 min), Day 2 (1441-2448 min) and Day 5 (7200-8640 min) in Band 880-915 MHz, and four different values of threshold: $\gamma=[-102, -104, -106, -108]$ dBm.  The parameters $U_{oc}$ and $L_{oc}$ will be selected by $M_s$, which represents the occupancy split that divides the data into occupied and idle classes. It varies from 0.1 to 0.9 with a step size of 0.1. It is observed in Fig.\ref{f11} that the value of CA depends on day and the value of threshold. The actual value of $OC^i_{train}$ in (\ref{e4}) always lies in a certain range, $[L_{s},U_{s}]$,  where $L_s$ represents the lowest value of $OC^i_{train}$ and $U_s$ represents the maximum value of $OC^i_{train}$. When $L_{s}<=M_s<=U_{s}$, two groups of classes $P^i=0$ (available class) and $P^i=1$ (occupied class) can be classified correctly. When $M_s>U_{s}$ or $M_s<L_{s}$, all the samples will be classified as one class because $OC^i_{train}$ is a closed set whose values do not lie outside the range $[L_{s},U_{s}]$. This explains why the $CA=1$ for $[L_{oc},U_{oc}]=[0.1,0.2]$ and $[L_{oc},U_{oc}]=[0.75,0.9]$ while $CA < 1$ for $[L_{oc},U_{oc}]= [0.2, 0.75]$ for Day 1 using $\gamma=-102$ dBm. Thus, the classification cannot be performed when $M_s>U_{s}$ or $M_s<L_{s}$. The optimal range is $[L_{oc},U_{oc}]=[0.2,0.75]$ for $CA<1$. However, for $CA\hspace{1mm}<\hspace{1mm}1$, there are four different choices of threshold available. In our proposed approach, we choose  that specific value of threshold that contains the largest number of values between $L_{oc}$ and $U_{oc}$. Following this, we have selected $\gamma=-102$ dBm for Day1, Day2 and Day5 as the optimal threshold which ensures the largest amount of samples between $L_{oc}$ and $U_{oc}$. The $[L_{oc},U_{oc}]=[0.2,0.75]$ for Day 1, $[L_{oc},U_{oc}]=[0.4,0.85]$ for Day2 and $[L_{oc},U_{oc}]=[0.2,0.80]$ for Day 5 respectively.  The optimal values of $\gamma$, $U_{oc}$ and $L_{oc}$ are further used for finding $B$ for each day.
\subsection{Model Performance Comparison}
Following the discussion above, we have compared the performance of the algorithms in this section using 1 month data of Band 880-915 MHz. Our tests show that the number of minimum observations/node for DT can be seclected as 17, number of predictors for LR as 15, normal kernel for NBC and linear kernel for SVM. The optimal splitting range, optimal threshold and $B$ will be selected corresponding to the data of each day. 

\subsubsection{Supervised VS Unsupervised Algorithms using $k=55$}
 In Fig. \ref{f16}(a), it is observed that the mean CA attained by LR, SVM, DT, NBC and HMM is 0.9257, 0.9162, 0.8483, 0.9493 and 0.4790 respectively. The mean computation time in each iteration by LR, SVM, DT, NBC and HMM is 350.19, 0.092, 0.0136, 0.0045, and 0.0171 seconds, respectively. Thus, NBC is the best considering the accuracy and complexity.
\subsubsection{Supervised vs Unsupervised Algorithms using $K=192$ }
\begin{figure}   
 
  \subfigure[\label{a}]{\includegraphics[width=4in]{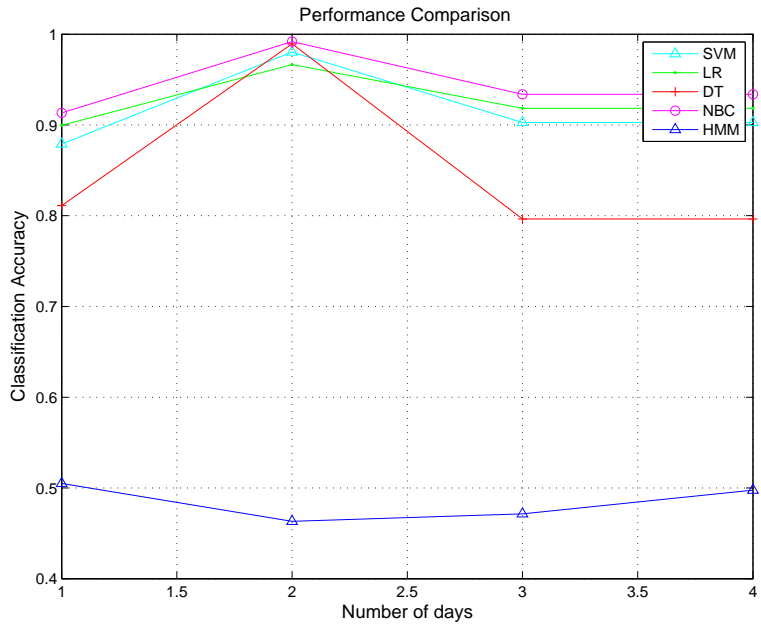}}

  \subfigure[\label{b}]{\includegraphics[width=4in]{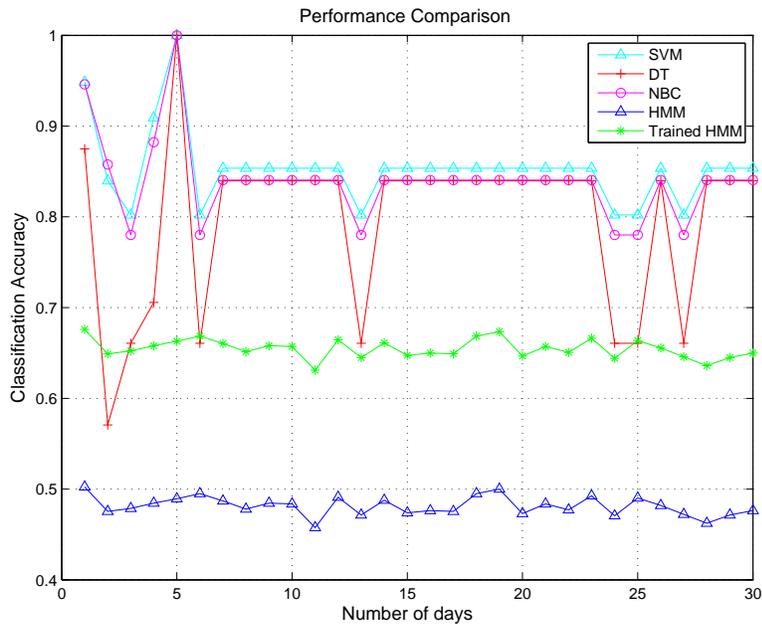}}

   \caption{Performance Comparison of (a) SVM, DT, NBC, LR and HMM with $k=55$. (b) SVM, DT, NBC, HMM and trained HMM with $k=192$.}
\label{f16}
\end{figure}

We have compared HMM, Trained HMM, SVM, DT and NBC in Fig.\ref{f16}(b) for 30 days. Each iteration represents 1 day.  LR is not shown as it takes an excessively long time in this case. It is observed that trained HMM performed better than HMM, but worst than DT, NBC and SVM.  The mean CA attained by Trained HMM, HMM, SVM, DT and NBC is 0.6816, 0.4887, 0.8528, 0.8392, 0.7970 while the computational time for each iteration of Trained HMM, HMM, SVM, DT and NBC 0.0205, 0.09066, 0.0135, 0.0163, 0.0095 seconds, respectively. Thus, SVM is the best in this case with highest CA and shortest time.
\subsubsection{SVM with Fire Fly Algorithm }
So far, the best overall performance is attained by the linear SVM technique. The performance of linear SVM is affected by the  value of $Box_{ct}$ as illustrated in Section IV-C. The fire fly algorithm can be used to select the best value of $Box_{ct}$. We set $\alpha=1$, $\beta_0=2$ and $\psi_{l_{1}l_{2}}=1.3$ for FFA. Fig. \ref{f17}(a) depicts that 'SVM+FFA' performs better than the conventional SVM in most of the cases.  The mean CA attained by SVM+FFA, SVM, DT, NBC and HMM is 0.8728, 0.8499, 0.7970, 0.8392 and 0.4822, respectively.
\subsubsection{Probability of SU Outage}
\begin{figure}

    \subfigure[\label{a}]{\includegraphics[width=4in]{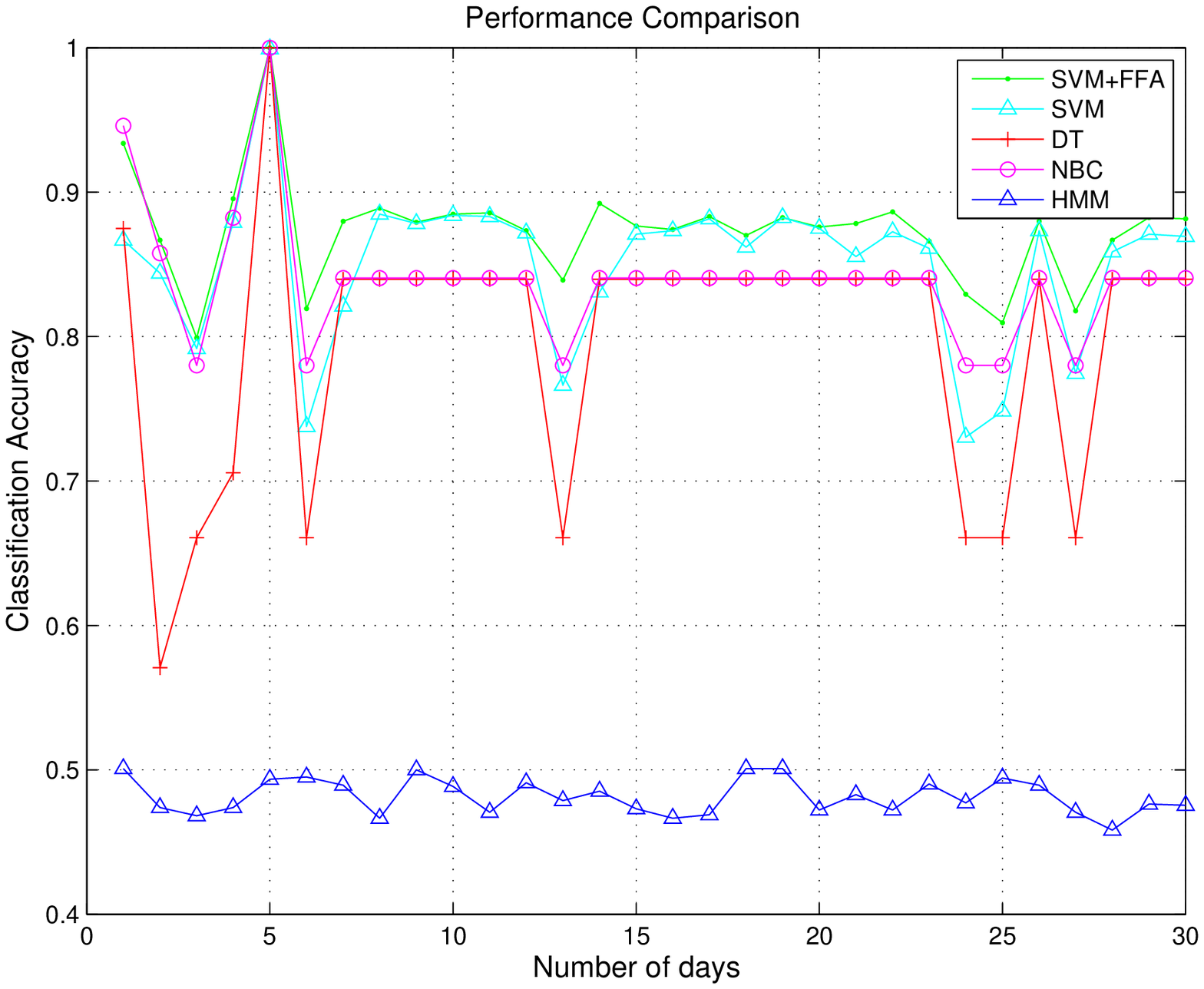}}

    \subfigure[\label{b}]{\includegraphics[width=4in]{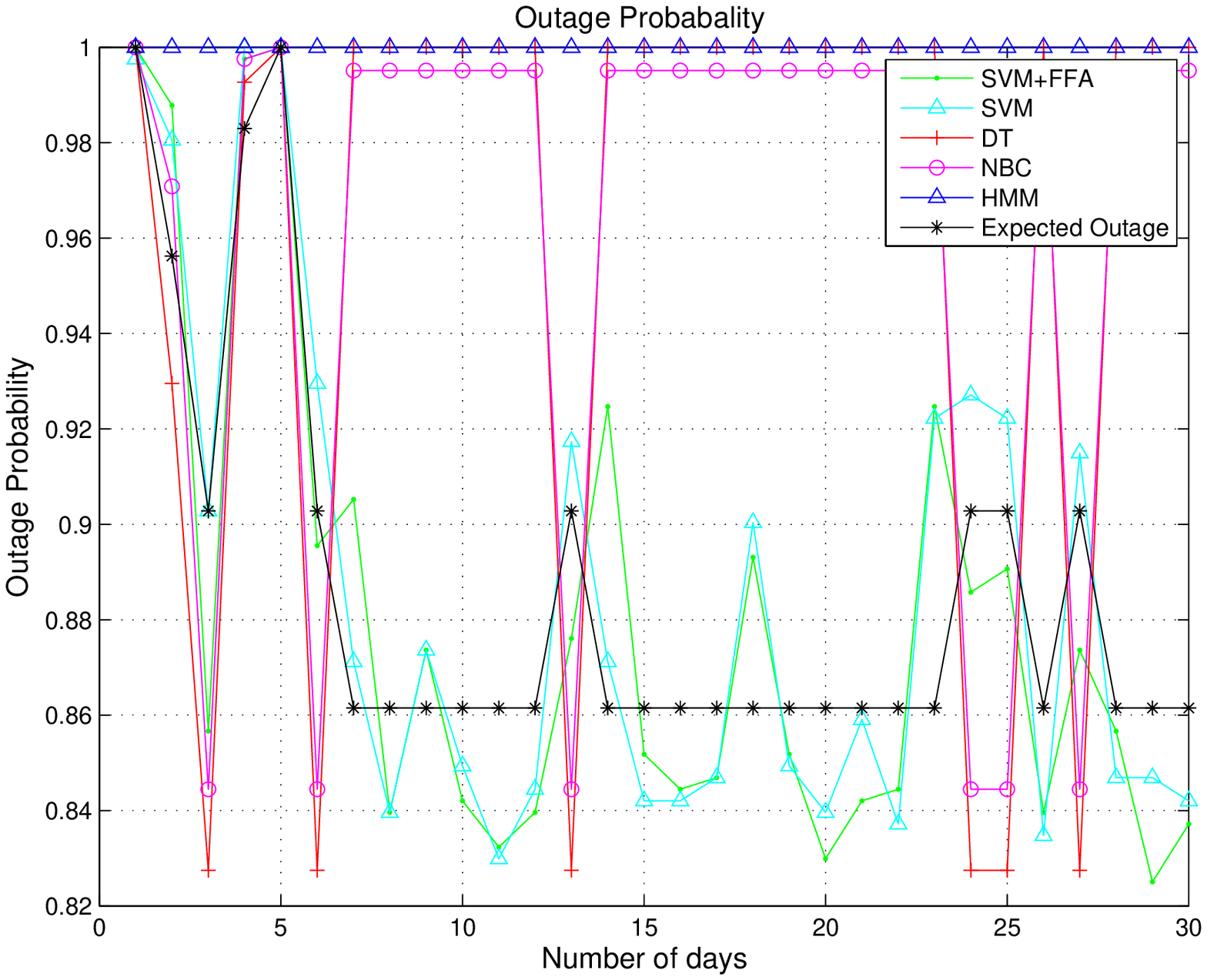}}

   \caption{Performance Comparison of ML algorithms: SVM, DT, NBC, HMM and 'SVM+FFA' using $k=192$ for a set of 30 days. (b) Comparison of 'expected probability of SU outage' with the SU outage evaluated using SVM, DT, NBC, HMM and 'SVM+FFA' using $k=192$ for a set of 30 days. }
\label{f17}
\end{figure}

This probability is computed using SVM+FFA, SVM, DT, NBC and HMM and compared with the expected $P(SU_{outage})$ to compute the difference between evaluated and expected values. It is evident in Fig. \ref{f17}(b) that SVM+FFA has predicted the $P(SU_{outage})$ with minimum difference and is very close to the expected one. The expected SU outage is 0.9191 in Fig. \ref{f17}(b) while the predicted $P(SU_{outage})$ using SVM+FFA, SVM, NBC, DT and HMM is 0.9264, 0.9322, 0.9638, 0.9577 and 1, respectively. The $P(SU_{outage})$ for HMM is always 1, which implies that HMM has failed to find any block of consecutive free time slot of length $out_{su}$.
\subsubsection{Supervised vs Unsupervised Algorithms using different Training/ Testing Data vectors}
We have presented the detailed comparison of supervised and unsupervised algorithms using different sizes of training and testing data Table 1. The classification accuracy and computation time for all supervised algorithms increases with an increase in the size of the training data.  SVM+FFA has attained the highest CA but with the longest computation time in most cases. 


\begin{table}[htbp]

\begin{center}

\begin{tabular}{|c|c|c|c|}\hline

\multicolumn{4}{ |c| }{Performance Comparison}

\\\hline

{Training data, Testing} data & Technique & Mean CA & Mean Computational Time \\\hline\hline

\multirow{5}{*}{{15 $\%$, 85 $\%$ }} 

& Decision Trees & 0.7612 & 0.0132 \\
& Support Vector Machine (SVM)&  0.8945 & 0.0128\\

& SVM + Fire Fly Algorithm & 0.9034 &  3.0412 \\

& Hidden Markov Model & 0.4925 &  0.0241 \\

& Naive Bayesian &  0.8714 & 0.0084\\

\hline \hline

%
%
%
%

\multirow{5}{*}{{30 $\%$, 70 $\%$ }} 

& Decision Trees & 0.8028 & 0.0198 \\
& Support Vector Machine (SVM) &  0.9143 & 0.0153\\

& SVM + Fire Fly Algorithm & 0.9189 &  3.8947\\

& Hidden Markov Model & 0.4841 &  0.0191 \\
& Naive Bayesian &  0.9064 & 0.0098\\

\hline \hline 
  \end{tabular}

%
%
%
%
%
%
%
%
%
%
%
%
%
%
%
%
%
%
%
%
 \caption{Performance Comparison of Five ML algorithms using different sizes of Training/Testig data.}

\label{tmu}

 \end{center}

 \end{table}

%
%
%

%
%
%
%

%
%
%
%

%
%
%
%
%
%
%
%

%

\newpage

\newpage

%
%
%
%
%
%

%
%
%
%
%
%
%


%
%
%
%
%
%

%
%
%
%
%
%


\begin{thebibliography}{10}

\bibitem{chen}   Y. Chen, H-S. Oh, "A survey of Measurement-based spectrum occupancy modelling for cognitive radios", \textit{ IEEE Communications Surveys and Tutorials}, Vol. PP, Issue. 99, pp. 1, Oct 2014.
%

%
%
\bibitem{occupation} V. Blaschke, H. Jaekel, T. Renk, C. Kloeck, F. K.  Jondral, "Occupation measurements supporting dynamic spectrum allocation for cognitive radio design", \emph{Proc. CrownCom'07}, pp. 50-57, Orlando, Florida, Aug. 2007. 




%
%
\bibitem{radio} S. Kaneko, S.  Nomoto, T. Ueda, S. Nomura and K. Takeuchi, " Predicting radio resource availability in cognitive radio - an experimental examination", \emph{CrownCom’08}, Singapore, May 2008.

\bibitem{predictive} M. Hoyhtya, S. Pollin, A. Mammela, "Classification-based predictive channel selection for cognitive radios", \emph{Proc. ICC’10}, pp. 1-6, Cape town, South Africa, May. 2010

\bibitem{power} X. Zhou, J. Ma, Y. Li, Y. H. Kwon, A. C. K. Soong,  G. Zhao, "Probability-based transmit power control for dynamic spectrum access", \emph{Proc. DySPAN’08}, pp. 1-5, Chicago, USA, Oct.  2008. 
%
\bibitem{duration}X. Zhou, J. Ma, Y. Li, Y. H. Kwon, A.C.K. Soong, "Probability-based optimization of inter-sensing duration and power control in cognitive radio", \emph{IEEE Transactions on Wireless Communications}, vol. 8, pp. 4922 - 4927, Oct. 2009.

%

%
%


\bibitem{statistics} Z. Wang, S. Salous, "Spectrum occupancy statistics and time series models for cognitive radio", \emph{ Journal of Signal Processing Systems}, vol.  62, Feb. 2011.

\bibitem{ML} C. Rudin, K. L. Wagstaff, "Machine learning for science and society", \emph{ Springer Journal on Machine Learning}, Vol. 95, Issue. 1, pp. 1-9, Nov 2013.

\bibitem{ml1} K. W. Choi, E. Hossain, D. I. Kin, "Cooperative Spectrum Sensing Under a Random Geometric Primary User Network Model", \emph{IEEE Transaction on Wireless Communications}, Vol. 10, No. 6, June 2011.

\bibitem{ml2} K. M. Thilina, K. W. Choi, N. Saquib, and E. Hossain, "Machine Learning Techniques for Cooperative

\bibitem{review} D. Willkomm, S. Machiraju, J. Bolot, A. Wolisz, "Primary Users in Cellular Networks: A Large-scale
Measurement Study", \emph{3rd IEEE Symposium on New Frontiers in Dynamic Spectrum Access Networks,  DySPAN' 08}, pp. 1-11.

\bibitem{pred} V. K. Tumuluru, P. Wang, D. Niyato, "Channel status prediction for cognitive radio networks", \textit{ Wiley Wireless Communications and Mobile Computing}, Vol. 12, Issue. 10, pp. 862-874, July 2012.

\bibitem{linear} S. Pagadarai  and A. M. Wyglinski, "A linear mixed-effects model of wireless spectrum occupancy", \emph{EURASIP Journal on Wireless Communications and Networking 2010},  Aug. 2010




\bibitem{energy} Z. Xuping, P. Jianguo, "Energy-detection based spectrum sensing for cognitive radio", \emph{Proc. CCWMSN’07}, pp. 944 -947, Dec. 2007 

\bibitem{utility} A. J. Petain, "Maximizing the Utility of Radio Spectrum: Broadband Spectrum Measurements and Occupancy Model for Use by Cognitive Radio",  \emph{Ph.D. dissertation}, Georgia Institute of Technology, Atlanta,  GA, USA, 2005.
%


\bibitem{nbapple} H. Zhang, "The Optimality of Naive Bayes", available online at, \url{http://courses.ischool.berkeley.edu/i290-dm/s11/SECURE/Optimality-of-Naive_Bayes.pdf}.

\bibitem{trees} L. Rokach, O. Maimon, "Decesion Trees", \emph{Data Mining and Knowledge Discovery Handbook}, Springer Publisher, 2nd ed. 2010, XX, 1285.

\bibitem{puni} K. Puniyani,  "Logistic regression and SVMs", available online at, \url{http://www.slideshare.net/NYCPredictiveAnalytics/intro-to-classification-logistic-regression-svm}.


\bibitem{svmd} A. b. Hur, J. Weston, "A user's guide to support vector machines", \emph{Data Mining Techniques for the Life Sciences Methods in Molecular Biology}, vol. 609,  pp.  223-239, 2010.


\bibitem{fire}X. Yang, "Firefly algorithms for multimodal optimization", \emph{LNCS 5792}, pp. 169–178, 2009.

\bibitem{hmm} P. Blunsom, "Hidden markov models", Aug 2004, available online at, \url{http://digital.cs.usu.edu/~cyan/CS7960/hmm-tutorial.pdf}.

\bibitem{hmm1} D. Garrette, J. Baldridge, "Type-supervised hidden Markov models for part-of-speech tagging with incomplete tag dictionaries", \emph{EMNLP-CoNLL'12}, pp. 821-831, Stroudsburg, PA, USA, 2012.

\bibitem{wiki}Kernel (statistics), available online at, \url{http://en.wikipedia.org/wiki/Kernel_(statistics)}.


\end{thebibliography}
\end{document}